\documentclass[runningheads]{llncs}
\usepackage{graphicx}
\usepackage{multirow}
\usepackage{siunitx}
\newcommand{\RNum}[1]{\uppercase\expandafter{\romannumeral #1 \relax}}

\begin{document}
\title{Environmental Sound Classification Based on Multi-temporal Resolution Convolutional Neural Network Combining with Multi-level Features}
\titlerunning{Multi-temporal Resolution \& Multi-level Features}
%
\author{Boqing Zhu\inst{1} \and
Kele Xu\inst{1,2}  \and
Dezhi Wang\inst{3}\thanks{Corresponding author.} \and
Lilun Zhang\inst{3} \and \\
Bo Li\inst{4} \and
Yuxing Peng\inst{1}
}

\authorrunning{Boqing, Kele, Dezhi et al.}
%

\institute{
Science and Technology on Parallel and Distributed Laboratory, National University of Defense Technology, Changsha, China \\
\email{zhuboqing09@nudt.edu.cn, pengyuxing@aliyun.com}\\
\and
School of Information Communication, National University of Defense Technology, Wuhan, China \\
\email{kelele.xu@Gmail.com}\\
\and
College of Meteorology and Oceanography, National University of Defense Technology, Changsha, China \\
\email{wang\_dezhi@hotmail.com, zll0434@163.com}
\and
Beijing University of Posts and Telecommunications, Beijing, China \\
\email{deepblue.lb@gmail.com}
}
%
\maketitle              
\begin{abstract}
Motivated by the fact that characteristics of different sound classes are highly diverse in different temporal scales and hierarchical levels, a novel deep convolutional neural network (CNN) architecture is proposed for the environmental sound classification task. This network architecture takes raw waveforms as input, and a set of separated parallel CNNs are utilized with different convolutional filter sizes and strides, in order to learn feature representations with multi-temporal resolutions. On the other hand, the proposed architecture also aggregates hierarchical features from multi-level CNN layers for classification using direct connections between convolutional layers, which is beyond the typical single-level CNN features employed by the majority of previous studies. This network architecture also improves the flow of information and avoids vanishing gradient problem. The combination of multi-level features boosts the classification performance significantly. Comparative experiments are conducted on two datasets: the environmental sound classification dataset (ESC-50), and DCASE 2017 audio scene classification dataset. Results demonstrate that the proposed method is highly effective in the classification tasks by employing multi-temporal resolution and multi-level features, and it outperforms the previous methods which only account for single-level features.

\end{abstract}
\section{Introduction}
\label{sec:introduction}
Audio classification aims to predict the most descriptive audio tags from a set of given tags determined before the analysis. Generally, it can be divided into three main sub-domains: environmental sound classification, music classification and speech classification. Environmental sound signals are quite informative in characterizing environmental context in order to achieve a detailed understanding of the acoustic scene itself \cite{marchi2016pairwise,mesaros2017dcase,stowell2015detection}. And a wide range of applications can be found in \cite{bugalho2009detecting,eyben2013recent}. Environmental sound classification (ESC) is also very important for machines to understand the surroundings, but it is still a challenging problem, which has attracted extensive interest recently. In particular, the deep-learning based methods using more complex neural networks \cite{han2016acoustic,fonseca2017acoustic,abesseracoustic} have shown great potential and significant improvement in this field. Due to the capability of learning hierarchical features from high-dimensional raw data, convolutional neural networks (CNNs) based approaches have become a choice in audio classification problem.

Time-frequency representation and its variants, such as spectrograms, mel-frequency cepstral coefficients (MFCCs) \cite{valero2012gammatone,boddapati2017classifying}, mel-filterbank features \cite{jung4dnn,deng2016university}, are the most popular input for CNN-based architectures. However, the hyper-parameters (such as hop size or window size) of short time Fourier transform (STFT) in the generation of these spectrogram-based representations is normally not particularly optimized for the task, while environmental sounds actually have different discriminative patterns in terms of time-scales and feature hierarchy \cite{lee2017sample,xu2016hierarchical,lee2017multi}. To avoid exhausting parameter search, this issue may be addressed by applying feature extraction networks that directly take raw audio waveforms as input. There are a decent number of CNN architectures that learns from raw waveforms \cite{sainath2015learning,dai2017very}. The majority of them employed large-sized filters in the input convolutional layers with various sizes of stride to capture frequency-selective responses, which are carefully designed to handle their target problems. There are also a few works that used small filter and stride sizes in the input convolution layers \cite{Tokozume2017Learning,palaz2015analysis} inspired by the VGG networks in image classification that use very small filters for convolutional layers.

Inspired by the fact that the different environmental sound tags have different performance sensitivity to different time-scales, a multi-scale convolutional neural network named WaveMsNet~\cite{zhu2018learning} was proposed to extract features by filter banks at multiple scales. It uses the waveform as input and facilitates learning more precise representations on a suitable temporal scale to discriminate difference of environmental sounds. After combining the representations of the different temporal resolutions, the proposed method claimed that superior performance can be achieved with waveform as input on the environmental sound classification datasets ESC-10 and ESC-50 \cite{ESC_Dataset}. Unlike previous attempts focusing on the adjustment the CNN architectures to enable the feature extraction at multi time-scales; in this paper, we explore to extend the approach to handle the multi-level features from hierarchical CNN layers together with multi-scale audio features in order to even further improve the current performance on the ESC problem. Similar to the network setup of DenseNet \cite{DenseNet}, the concatenation of multi-level features is implemented by direct connections between convolutional layers in a feed-forward fashion, which accounts for more hierarchical features and also allows convolutional networks to be more efficient to train with the help of similar mechanism of skip connections \cite{ResNet}. Moreover, our method is also evaluated on another benchmark dataset from the DCASE 2017 audio scene classification task to demonstrate the generalization capability.

In this study, we have following contributions: 1) a novel CNN-based architecture is designed that is capable of comprehensively combining the audio features with multi-temporal resolutions from raw waveforms and the multi-level features from different CNN hierarchical layers. 2) Comparatively studies are conducted to demonstrate the effect of multi time-scale and multi-level features on the classification performance of environmental sounds. 3) Explore to visualize the learned multi-temporal resolution and multi-level audio features to explain the physical meaning of what the model has really learned.

The rest of the paper is organized as follows. Section 2 discusses related work. In Section 3, we describe our proposed architecture with implementation details. The experimental setup and results are given in section 4, while Section 5 concludes this paper.

\section{Related Work} 
\label{sec:related_work}

Due to the rapid development in signal processing and machine learning domains, there is an extensive surge of interest in applying deep learning approaches for the audio classification (or audio tagging) task. Most of the approaches with good performance \cite{jung4dnn,han2017convolutional} for the environmental sound classification related tasks of the DCASE 2017 challenge \cite{mesaros2017dcase} utilize deep learning models such as CNNs, which have already become the most popular method. The frequency based features are commonly used as input of CNN models in the environmental sound classification. The frequency based features are also replaced with raw audio waves as the input for the classifiers in some studies. This kind of end-to-end learning approach has been successfully used in speech recognition \cite{sainath2015learning}, music genre recognition \cite{dieleman2014end} and so on. Recently, a raw waveform-based approach so-called Sample-CNN model \cite{lee2018samplecnn} shows comparable performance to the spectrogram-based CNN model in music tagging by using sample-level filters to learn hierarchical audio characteristics.

Most of the previous studies in environmental sound classification utilize only one level or one scale of features for the classification, which is typically adopted in image classification. However, this kind of method ignores that, for the audio, discriminative features are generally positioned in different levels or time-scales in a hierarchy. This issue is addressed in some work by comparing or combining multi-layer or multi-scale audio features \cite{hamel2012building,dieleman2013multiscale}. The combination of different resolutions of spectrograms in terms of time-scale \cite{hamel2012building} is extensively studied for the prediction of audio tags. This idea is also further improved by using Gaussian and Laplacian pyramids \cite{dieleman2013multiscale}. Instead of concatenating the multi-scaling features only on the input layer, attempts are also made to combine audio features from different levels \cite{lee2017multi}, which is believed to provide a superior performance. To increase temporal resolutions of Mel-spectrogram segments for acoustic scene classification, an architecture \cite{schindlermulti} consisting of parallel convolutional neural networks is presented where it shows a significant improvement compared with the  best single resolution model. In the paper \cite{xu2018mixup}, mixup method is explored to provide higher prediction accuracy and robustness.

\begin{figure}[t!]
\includegraphics[width=\textwidth]{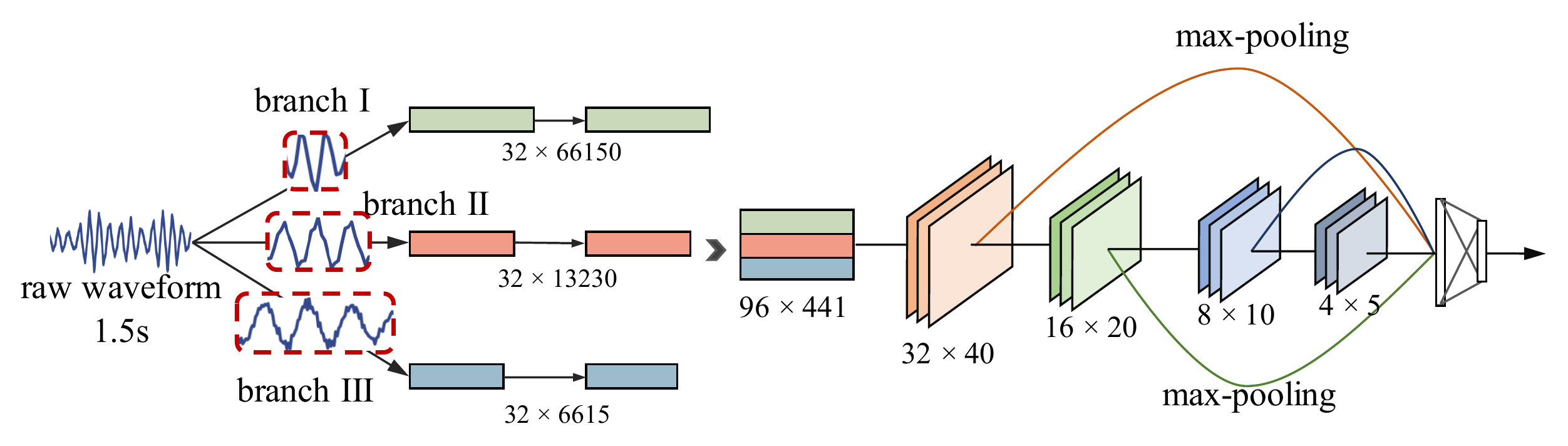}
\caption{Network architecture for environmental sound classification.}
\label{network}
\end{figure}

\section{Proposed Method}
In this section, we investigate the combination of multi-temporal resolution and multi-level features, for the problem of environmental sound classification.

\subsection{Overview}
The proposed network architecture is presented in the Fig.\ref{network}. The network is designed as an end-to-end system which takes the wave signal as input and class label as output. When training the network, we randomly select 1.5 seconds from the original training raw waveform data and input it into the network. The selected section is different in each epoch, and we use the same training label regardless of the selected section. When testing, we classify testing data based on probability-voting. That is, we divide the testing audio into multiple 1.5s sections and input each of them into the network. We take the sum of all the output probabilities after softmax and use it to classify the testing data.

\subsection{Multi-temporal Resolution CNN} 
The architecture is composed of a set of separated parallel 1-D time domain convolutional layers with different filter sizes and strides, in order to learn feature representations with multi-temporal resolutions. Specifically, to learn high-frequency features, filters with a short window are applied at a small stride. Low-frequency features, on the contrary, employ a long window that can be applied at a larger stride. Then feature maps with different temporal resolution are concatenated along frequency axis and pooled to the same dimension on the time axis.

In our experiments, we apply three branches of separated parallel 1-D convolutional layers (branch \RNum{1}: (size 11, stride 1), branch \RNum{2}: (size 51, stride 5), branch \RNum{3}: (size 101, stride 10)). Each branch has 32 filters. Another time-domain convolutional layer is followed to create invariance to phase shifts with filter size 3 and stride 1. We aggressively reduce the temporal resolution to 441 with a max pooling layer to each branch ’s feature map. Then we concatenate three feature map together to get the multi-temporal resolution features to represent the audios.

\subsection{Multi-level Feature Concatenation} 
Next, we apply four convolutional layers for the multi-temporal resolution feature map. The two dimensions of the feature map correspond to $frequency \times time$. There are 64, 128, 256, 256 filters in each convolutional layer respectively with a size of 3 $\times$ 3, and we stride the filter by 1 $\times$ 1. We leverage non-overlapping max pooling to down-sample the features to the corresponding size as shown in Fig.\ref{network}. The outputs of the four convolutional layers are concatenated and then delivered to the full connection layers. Before the concatenation, the dimensions of the outputs are reduced to 4 $\times$ 5 by max pooling. In the experimental section, we investigate the effect of the concatenated layers in multi-level features. The input size of full connection layer adjusts to the dimensionality of the concatenated feature maps. For instance, when we pick features from last 3 layers, the model will have $(128 + 256 + 256) \times 4 \times 5$ dimensional feature maps.

\section{Experiments}
In this section, details of the DCASE 2017 ASC dataset and ESC-50 dataset used in the experiment are first introduced. Then the model parameters and experimental setup are presented for the comparison of performance between the proposed model and the previous models. We performed 5-fold cross-validation five times on the dataset. Finally, the conclusion is drawn based on the experimental results.

\subsection{Dataset}
We use two datasets: 2017 DCASE challenge dataset for audio scene classification task and ESC-50 dataset to validate the performance of proposed method. DCASE challenge dataset~\cite{mesaros2017dcase,stowell2015detection} is established to determine the context of a given recording through selecting one appropriate label from a pre-determined set of 15 acoustic scenes such as cafe/restaurant, car, city center and so on. Each scene contains 312 recordings with a length of 10 seconds, a sampling rate of 44.1 kHz and 24-bit resolution in stereo in the development dataset. Totally there are 4680 audio recordings in the development dataset which is provided at the beginning of the challenge, together with ground truth. Besides, an evaluation dataset is also released with 1620 audio recordings in total after the challenge submission is closed. A four-fold cross-validation setup is provided so as to make results reported strictly comparable. The evaluation dataset is used to evaluate the performance of classification models.

ESC-50~\cite{ESC_Dataset} dataset which is public labeled sets of environmental recordings are also used in our experiments. ESC-50 dataset comprises 50 equally balanced classes, each clip is about 5 seconds and sampled at 44.1kHz. The 50 classes can be divided into 5 major groups: animals, natural soundscapes and water sounds, human non-speech sound, interior/domestic sounds, and exterior/urban noises. Datasets have been prearranged into 5 folds for comparable cross-validation and other experiments~\cite{Tokozume2017Learning} used these folds. The same fold division is employed in our evaluation. The metric used is classification accuracy, and the average accuracy across the five folds is reported for comparison.

\subsection{Experimental Details}
For the network training, cross-entropy loss is used. To optimize the loss, the momentum stochastic gradient descent algorithm is applied with momentum 0.9. We use Rectified Linear Units (ReLUs) to implement nonlinear activation functions. A batch size of 64 is applied. All weight parameters are subjected to $\ell_2$ regularization with coefficient 5 $\times$ $\num{e-4}$. We train models for 160 epochs until convergence. Learning rate is set as $\num{e-2}$ for first 60 epochs, $\num{e-3}$ for next 60 epochs, $\num{e-4}$ for next 20 epochs and $\num{e-5}$ for last 20 epochs. The weights in the time-domain convolutional layers are randomly initialized. The models in experiment are implemented by PyTorch~\cite{paszke2017automatic} and trained on GTX Titan X GPU cards. We randomly select a 1.5 seconds waveform as input when training the model. In testing phase, we use the probability-voting strategy.

\begin{table}[t]
\renewcommand\arraystretch{1.2}
\centering
\caption{Comparison of Multi-Temporal Resolution and Single-Temporal Resolution.}
\label{tb:multi_temporal}
\begin{tabular}{cccccc}
\noalign{\smallskip}\hline\noalign{\smallskip}
\multirow{2}{*}{\begin{tabular}[c]{@{}c@{}}Temporal\\ Resolution\end{tabular}} &
\multicolumn{3}{c}{Filter Number} & \multicolumn{2}{c}{Mean Accuracy (\%)}   \\

        & branch \RNum{1}  & branch \RNum{2}  & branch \RNum{3}  & ESC-50           & DCASE 2017          \\
\noalign{\smallskip}\hline\noalign{\smallskip}
Low     & 96  & 0   & 0  & 69.1 $\pm$ 2.63  & 70.3 $\pm$ 3.63     \\
Middle  & 0   & 96  & 0  & 68.2 $\pm$ 2.29  & 71.6 $\pm$ 3.78     \\
High    & 0   & 0   & 96 & 68.4 $\pm$ 3.13  & 71.3 $\pm$ 4.02     \\
Multi   & 32  & 32  & 32 & \textbf{71.6} $\pm$ 2.58  &  \textbf{73.1} $\pm$ 3.34\\
\hline
baseline \cite{Piczak2015Environmental,mesaros2017dcase} & \multicolumn{3}{c}{-}             & 64.5  &     61.0   \\
\noalign{\smallskip}\hline\noalign{\smallskip}
\end{tabular}
\end{table}

\begin{table}[t]
\renewcommand\arraystretch{1.2}
\centering
\caption{Performance with Multi-Level Feature.}
\label{tb:multi_level}
\begin{tabular}{cccccc}
\noalign{\smallskip}\hline\noalign{\smallskip}
\multirow{2}{*}{Last $N$ layers feature} &
\multicolumn{2}{c}{Mean Accuracy (\%)}   \\

          & ESC-50           & DCASE 2017          \\
\noalign{\smallskip}\hline\noalign{\smallskip}
$N = 1$        & 71.6 $\pm$ 2.58  & 73.1 $\pm$ 3.34 \\
$N = 2$        & 71.8 $\pm$ 2.79  & 73.2 $\pm$ 3.27 \\
$N = 3$        & 73.0 $\pm$ 2.19  & 73.9 $\pm$ 2.95 \\
$N = 4$        & \textbf{73.2} $\pm$ 2.90  & \textbf{74.7} $\pm$ 2.46 \\
\noalign{\smallskip}\hline\noalign{\smallskip}
\end{tabular}
\end{table}

\subsection{Results}

\subsubsection{Effect of multi-temporal resolution.}
We compare the performances with constant filter size at three different temporal resolutions, low temporal-resolution (Low), middle temporal-resolution (Middle) and high temporal-resolution (High). These three models remain only one corresponding branch (Low remains branch \RNum{1}, Middle remains branch \RNum{2} and High remains branch \RNum{3}). As we reduce the number of convolution filters, it may cause performance degradation. So we use triple filters in time-domain convolution in single temporal resolution models for fair comparison. These three variant models are trained separately. Table \ref{tb:multi_temporal} demonstrates the mean accuracy and standard error using multi-temporal resolution features and single-temporal resolution features. Both single-temporal resolution CNNs and multi-temporal resolution CNN showed better performance against the baseline. Our multi-temporal resolution model achieves average improvement of 3.0\% and 2.0\% compared with the single-temporal resolution models on ESC-50 and DCASE2017 dataset respectively.

\subsubsection{Effect of multi-level features.}
Next, we demonstrate the effectiveness of multiple-level features. We down-sample and stack the feature map of the last $N$ ($N=1, 2, 3, 4$) layers of the network. They are delivered to the full connection layers. When $N=1$, single-level features are used. As demonstrated in Table \ref{tb:multi_level}, the accuracies consistently increase on both datasets. Further, the performances are always benefited from the increase of $N$. When $N=4$, that is, concatenating features of each layer of 2D convolution layer, we got the best result on ESC-50 and DCASE2017.

\begin{figure}[t!]
\includegraphics[width=\textwidth]{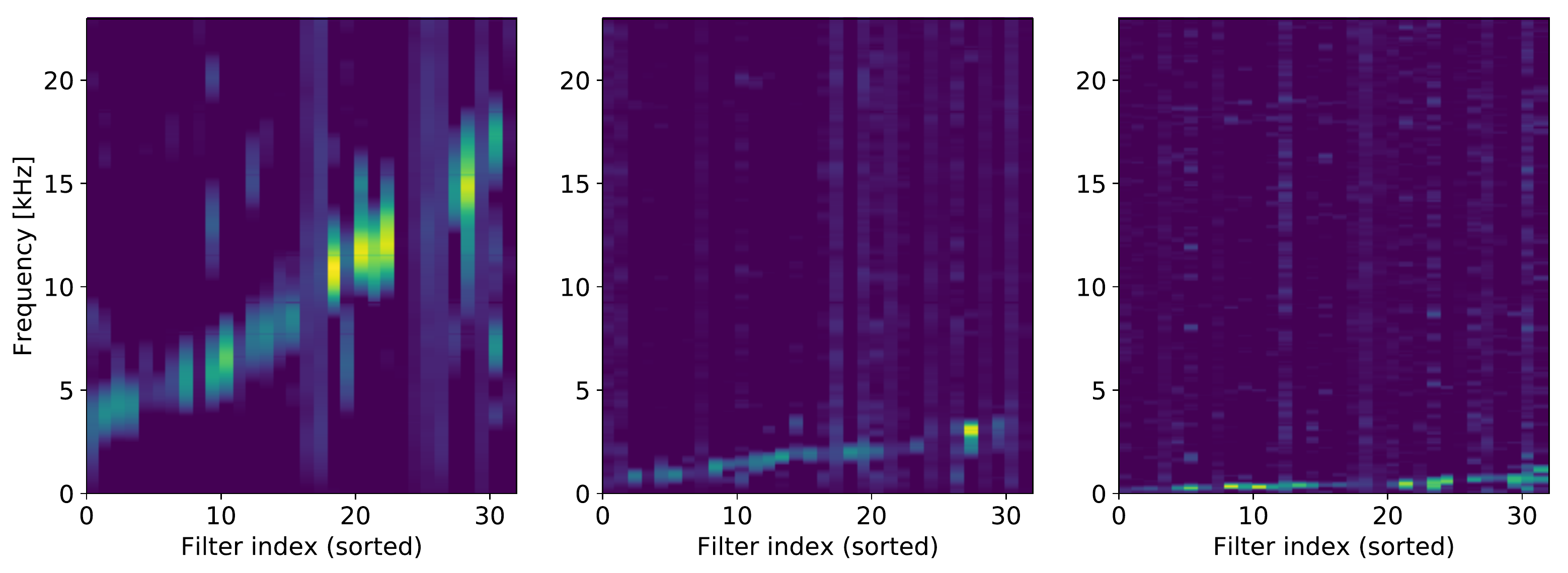}
\caption{Frequency response of the multi-temporal resolution feature maps. Left shows the frequency response of feature map product by branch \RNum{1}. Middle corresponds to branch \RNum{2}. Right corresponds to branch \RNum{3}.}
\label{responses}
\end{figure}

\subsubsection{Analysis of the results.}
Here, we present the analysis of the multi-temporal resolution features. The technique of visualizing the filters at different branches~\cite{lee2018samplecnn} can provide deeper understanding of what the networks have learned from the raw waveforms. Fig.\ref{responses} shows the responses of the multi-temporal resolution feature maps. Most of the filters learn to be band-pass filters while the filters are sorted by their central frequencies from low to high as shown in the figure. Branch \RNum{1} has learned more dispersed bands across the frequency that can extract the features from all frequencies. But the frequency resolution is lower. On the contrary, branch III has learned high-frequency resolution bands and most of them locate at the low-frequency area. Branch \RNum{2} behaves between branch \RNum{1} and \RNum{3}. This indicts that different branch could learn discrepant features, and the filter banks split responsibilities based on what they efficiently can represent. This explains why multi-temporal resolution models get a better performance than the single-temporal resolution model shown in Table \ref{tb:multi_temporal}.

\section{Conclusion} 
\label{sec:conclusion}
In this article, we proposed an effective CNN architecture integrating the networks for multi-temporal resolution analysis and multi-level feature extraction in order to achieve more comprehensive feature representations of audios and tackle the multi-scale problem in the environmental sound classification. Through the experiments, it is shown that combining the multi-level and multi-scale features improves the overall performance. The raw waveforms are directly taken as the model input, which enables the proposed approach to be applied in an end-to-end manner. The frequency response of learned filters at the model branches with different temporal resolutions is visualized to better interpret the multi time-scale effect on filter characteristics. In future, we would like to evaluate the performance of our method on a large-scale dataset of Google AudioSet for the general-purpose audio tagging task.

\section*{Acknowledgments}
This study was funded by the National Key R\&D Program of China under Grant No.2016YFC1401800, the National Natural Science Foundation of China (No.61379056, No.61702531) and the Scientific Research Project of NUDT (No.ZK\\17-03-31, No.ZK16-03-46).

%
%
%
\bibliographystyle{splncs04}
\bibliography{bib_ref}

\begin{thebibliography}{10}
\providecommand{\url}[1]{\texttt{#1}}
\providecommand{\urlprefix}{URL }
\providecommand{\doi}[1]{https://doi.org/#1}

\bibitem{abesseracoustic}
Abe{\ss}er, J., Mimilakis, S.I., Gr{\"a}fe, R., Lukashevich, H., Fraunhofer,
  I.: Acoustic scene classification by combining autoencoder-based
  dimensionality reduction and convolutional neural networks  (2017)

\bibitem{boddapati2017classifying}
Boddapati, V., Petef, A., Rasmusson, J., Lundberg, L.: Classifying
  environmental sounds using image recognition networks. Procedia Computer
  Science  \textbf{112},  2048--2056 (2017)

\bibitem{bugalho2009detecting}
Bugalho, M., Portelo, J., Trancoso, I., Pellegrini, T., Abad, A.: Detecting
  audio events for semantic video search. In: Tenth Annual Conference of the
  International Speech Communication Association (2009)

\bibitem{dai2017very}
Dai, W., Dai, C., Qu, S., Li, J., Das, S.: Very deep convolutional neural
  networks for raw waveforms. In: Acoustics, Speech and Signal Processing
  (ICASSP), 2017 IEEE International Conference on. pp. 421--425. IEEE (2017)

\bibitem{deng2016university}
Deng, J., Cummins, N., Han, J., Xu, X., Ren, Z., Pandit, V., Zhang, Z.,
  Schuller, B.: The university of passau open emotion recognition system for
  the multimodal emotion challenge. In: Chinese Conference on Pattern
  Recognition. pp. 652--666. Springer (2016)

\bibitem{dieleman2013multiscale}
Dieleman, S., Schrauwen, B.: Multiscale approaches to music audio feature
  learning. In: 14th International Society for Music Information Retrieval
  Conference (ISMIR-2013). pp. 116--121. Pontif{\'\i}cia Universidade
  Cat{\'o}lica do Paran{\'a} (2013)

\bibitem{dieleman2014end}
Dieleman, S., Schrauwen, B.: End-to-end learning for music audio. In:
  Acoustics, Speech and Signal Processing (ICASSP), 2014 IEEE International
  Conference on. pp. 6964--6968. IEEE (2014)

\bibitem{eyben2013recent}
Eyben, F., Weninger, F., Gross, F., Schuller, B.: Recent developments in
  opensmile, the munich open-source multimedia feature extractor. In:
  Proceedings of the 21st ACM international conference on Multimedia. pp.
  835--838. ACM (2013)

\bibitem{fonseca2017acoustic}
Fonseca, E., Gong, R., Bogdanov, D., Slizovskaia, O., G{\'o}mez~Guti{\'e}rrez,
  E., Serra, X.: Acoustic scene classification by ensembling gradient boosting
  machine and convolutional neural networks. In: Virtanen T, Mesaros A,
  Heittola T, Diment A, Vincent E, Benetos E, Martinez B, editors. Detection
  and Classification of Acoustic Scenes and Events 2017 Workshop (DCASE2017);
  2017 Nov 16; Munich, Germany. Tampere (Finland): Tampere University of
  Technology; 2017. p. 37-41. Tampere University of Technology (2017)

\bibitem{hamel2012building}
Hamel, P., Bengio, Y., Eck, D.: Building musically-relevant audio features
  through multiple timescale representations. In: ISMIR. pp. 553--558 (2012)

\bibitem{han2016acoustic}
Han, Y., Lee, K.: Acoustic scene classification using convolutional neural
  network and multiple-width frequency-delta data augmentation. arXiv preprint
  arXiv:1607.02383  (2016)

\bibitem{han2017convolutional}
Han, Y., Park, J., Lee, K.: Convolutional neural networks with binaural
  representations and background subtraction for acoustic scene classification
  (2017)

\bibitem{ResNet}
He, K., Zhang, X., Ren, S., Sun, J.: Deep residual learning for image
  recognition. In: Computer Vision and Pattern Recognition. pp. 770--778 (2016)

\bibitem{DenseNet}
Huang, G., Liu, Z., Weinberger, K.Q., van~der Maaten, L.: Densely connected
  convolutional networks. In: Proceedings of the IEEE conference on computer
  vision and pattern recognition. vol.~1, p.~3 (2017)

\bibitem{jung4dnn}
Jung, J.W., Heo, H.S., Yang, I.H., Yoon, S.H., Shim, H.J., Yu, H.J.: Dnn-based
  audio scene classification for dcase 2017: Dual input features, balancing
  cost, and stochastic data duplication. System  \textbf{4}, ~5

\bibitem{lee2017multi}
Lee, J., Nam, J.: Multi-level and multi-scale feature aggregation using
  pretrained convolutional neural networks for music auto-tagging. IEEE signal
  processing letters  \textbf{24}(8),  1208--1212 (2017)

\bibitem{lee2017sample}
Lee, J., Park, J., Kim, K.L., Nam, J.: Sample-level deep convolutional neural
  networks for music auto-tagging using raw waveforms. arXiv preprint
  arXiv:1703.01789  (2017)

\bibitem{lee2018samplecnn}
Lee, J., Park, J., Kim, K.L., Nam, J.: Samplecnn: End-to-end deep convolutional
  neural networks using very small filters for music classification. Applied
  Sciences  \textbf{8}(1), ~150 (2018)

\bibitem{marchi2016pairwise}
Marchi, E., Tonelli, D., Xu, X., Ringeval, F., Deng, J., Squartini, S.,
  Schuller, B.: Pairwise decomposition with deep neural networks and multiscale
  kernel subspace learning for acoustic scene classification. In: Proceedings
  of the Detection and Classification of Acoustic Scenes and Events 2016
  Workshop (DCASE2016). pp. 65--69 (2016)

\bibitem{mesaros2017dcase}
Mesaros, A., Heittola, T., Diment, A., Elizalde, B., Shah, A., Vincent, E.,
  Raj, B., Virtanen, T.: Dcase 2017 challenge setup: Tasks, datasets and
  baseline system. In: DCASE 2017-Workshop on Detection and Classification of
  Acoustic Scenes and Events (2017)

\bibitem{palaz2015analysis}
Palaz, D., Magimai.-Doss, M., Collobert, R.: Analysis of cnn-based speech
  recognition system using raw speech as input. Tech. rep., Idiap (2015)

\bibitem{paszke2017automatic}
Paszke, A., Gross, S., Chintala, S., Chanan, G., Yang, E., DeVito, Z., Lin, Z.,
  Desmaison, A., Antiga, L., Lerer, A.: Automatic differentiation in pytorch
  (2017)

\bibitem{ESC_Dataset}
Piczak, K.J.: {ESC}: {Dataset} for {Environmental Sound Classification}. In:
  Proceedings of the 23rd {Annual ACM Conference} on {Multimedia}. pp.
  1015--1018. {ACM Press}. \doi{10.1145/2733373.2806390},
  \url{http://dl.acm.org/citation.cfm?doid=2733373.2806390}

\bibitem{Piczak2015Environmental}
Piczak, K.J.: Environmental sound classification with convolutional neural
  networks. In: Machine Learning for Signal Processing (MLSP), 2015 IEEE 25th
  International Workshop on. pp.~1--6. IEEE (2015)

\bibitem{sainath2015learning}
Sainath, T.N., Weiss, R.J., Senior, A., Wilson, K.W., Vinyals, O.: Learning the
  speech front-end with raw waveform cldnns. In: Sixteenth Annual Conference of
  the International Speech Communication Association (2015)

\bibitem{schindlermulti}
Schindler, A., Lidy, T., Rauber, A.: Multi-temporal resolution convolutional
  neural networks for the dcase acoustic scene classification task

\bibitem{stowell2015detection}
Stowell, D., Giannoulis, D., Benetos, E., Lagrange, M., Plumbley, M.D.:
  Detection and classification of acoustic scenes and events. IEEE Transactions
  on Multimedia  \textbf{17}(10),  1733--1746 (2015)

\bibitem{Tokozume2017Learning}
Tokozume, Y., Harada, T.: Learning environmental sounds with end-to-end
  convolutional neural network. In: IEEE International Conference on Acoustics,
  Speech and Signal Processing. pp. 2721--2725 (2017)

\bibitem{valero2012gammatone}
Valero, X., Alias, F.: Gammatone cepstral coefficients: Biologically inspired
  features for non-speech audio classification. IEEE Transactions on Multimedia
   \textbf{14}(6),  1684--1689 (2012)

\bibitem{xu2018mixup}
Xu, K., Feng, D., Mi, H., Zhu, B., Wang, D., Zhang, L., Cai, H., Liu, S.:
  Mixup-based acoustic scene classification using multi-channel convolutional
  neural network. arXiv preprint arXiv:1805.07319  (2018)

\bibitem{xu2016hierarchical}
Xu, Y., Huang, Q., Wang, W., Plumbley, M.D.: Hierarchical learning for
  dnn-based acoustic scene classification. arXiv preprint arXiv:1607.03682
  (2016)

\bibitem{zhu2018learning}
Zhu, B., Wang, C., Liu, F., Lei, J., Lu, Z., Peng, Y.: Learning environmental
  sounds with multi-scale convolutional neural network. arXiv preprint
  arXiv:1803.10219  (2018)

\end{thebibliography}

\end{document}